\documentstyle[aps,12pt,manuscript]{revtex}
\begin{document}
\preprint{INJE--TP--96--2 }
\def\overlay#1#2{\setbox0=\hbox{#1}\setbox1=\hbox to \wd0{\hss #2\hss}#1%
\hskip -2\wd0\copy1}

\title{ New divergences of tachyon in the two-dimensional charged black holes}

\author{ H.W. Lee and Y. S. Myung }
\address{Department of Physics, Inje University, Kimhae 621-749, Korea} 
\author{Jin Young Kim}
\address{Division of Basic Science, Dongseo University, Pusan 616-010, Korea}
\author{D. K. Park}
\address{Department of Physics, Kyungnam University, Masan 631-701, Korea}
\maketitle
\vskip 1.5in

\begin{abstract}
We  quantize the tachyon field in the  two-dimensional (2D), $\epsilon<2$ 
charged black holes  where $\epsilon$ is the dilaton coupling parameter  for the Maxwell term.
Especially the expectation value of the stress-energy tensor $\langle T_{ab}\rangle$, 
observed by a freely falling observer, is computed.   This shows that
new divergences such as $\ln f$ and ${1 \over f}$ arises near the horizon ($f \to 0$), 
 compared with conformal matter case. 

\end{abstract}
\vskip .5in
PACS number(s) : 04.70.Dy, 04.60.Kz

\newpage
Lower dimensional theories of gravity provide  the simplified contexts in which to study
black hole physics [1]. The non-triviality of these models
arises from the non-minimal coupling of the dilaton to the scalar curvature. 
A dilaton potential of the type produced by the string loop corrections may induce multiple
horizons [2].   For example, 2D charged black hole from heterotic string theories has shown this 
feature. This has many analogies with  the Reissner-Nordstr\"om black hole in 4D general relativity. 

It is  important to investigate the classical stability of the black holes,
which  is essential to establish their
physical existence [3]. It has been shown that the 2D dilaton black hole is stable [4],
while  the extremal black holes  are shown to be classically  untable [5].
Furthermore the quantum stability is also important [6], since the back reaction effects due to
the non-zero stress-energy of the quantum field will change the background  geometry of space-time
near the  event horizon.  Even if it is classically stable, the instability may be caused by
the divergence 
of the renormalized expectation value of the the stress-energy tensor($\langle T_{\mu\nu}\rangle$)
 associated with
the quantized matter field. If one finds  a divergence of the stress-energy, the solution to
the quantum theory does not exist near  the classical solution and thus the quantum effects
 alter drastically the classical spacetime geometry [7]. As a  result, 
the  black hole is unstable quantum mechanically if there exists a divergence of 
the stress-energy. 
A conformally invariant scalar field ($f_i$) is usually used to study the quantum aspects
 of black hole [8].
If one takes a conformally invariant matter to study the classical aspects of the black
hole, one finds the free field equation for the perturbation : $ \nabla^2 f_i=0 \to
 (d^2/dr^{*2} +\omega^2)f_i=0$. This implies that one cannot find
the potential, which is crucial for obtaining information about the 2D black hole. 
Although $f_i$ is a simple matter field for the quantum study of
the black hole, it does not include  all information for the 2D black holes.

Here we introduce a tachyon as a test field. This provides us the potential that illustrates
many qualitative results about the 2D charged black holes.
Further new quantum results  are expected, because the tachyon  is coupled nontrivially 
to dilation.
In this paper, we consider the two-dimensional dilaton gravity coupled to
Maxwell and tachyon fields.
The relevant coupling (parametrized by  $\epsilon$) between the dilaton and Maxwell field 
is included to obtain the general 2D, $\epsilon<2$ charged black holes.
 This may be considered as a two-dimensional counterpart
of the 4D dilaton gravity with the parameter $a$ [9].

Trivedi [10] showed that the  stress-energy tensor of a conformal matter
diverges at the horizon in the 2D, $\epsilon=0$ extremal black hole.
This divergence can be better understood by regarding the extremal black hole as the limit of the
non-extremal one. The non-extremal black hole has both the outer (event) and inner (Cauchy) horizons,
 and the two horizons come together in the extremal limit.  In this case,
it is found that if one adjusts  the quantum state of the scalar field so that
the stress-energy tensor is finite at the outer horizon, it always diverges at the inner horizon.

We  start with  two-dimensional dilaton ($\Phi$) gravity  conformally coupled to Maxwell ($F_
{\mu\nu}$) and tachyon ($T$) fields [2,5,12]
\begin{equation}
S = {1 \over 2 \pi} \int d^2 x \sqrt{-G} e^{-2\Phi}
   \big \{ R + 4 (\nabla \Phi)^2 + 4\lambda^2 - {1 \over 2}e^{\epsilon \Phi} F^2 - 
 (\nabla T)^2  - V(T) \big \}
\end{equation}
with the tachyon potential $V(T)= -m_0^2 T^2$.
Our sign conventions and notation follow Misner, Thorne, and Wheeler [11].
The above action with $\epsilon=0$ can be realized from  the 
heterotic string.
Then the equations of motion become
\begin{eqnarray}
&&R_{\mu\nu} + 2\nabla_\mu \nabla_\nu \Phi  - \nabla_\mu T \nabla_\nu T 
-{ 4 - \epsilon \over 4}e^{\epsilon \Phi}  F_{\mu\rho}F_{\nu}^{~\rho} = 0,  \\
&& \nabla^2 \Phi -2 (\nabla \Phi)^2  + {1 \over 4}e^{\epsilon \Phi} F^2  + {m_0^2 \over2} T^2 +
 2\lambda^2  = 0,  \\
&&\nabla_\mu F^{\mu \nu} - (2- \epsilon) (\nabla_\mu \Phi) F^{\mu \nu} = 0,   \\
&&\nabla^2 T - 2 \nabla \Phi \cdot \nabla T + m_0^2 T = 0.
\end{eqnarray}

The general solution to (2)-(5) with tachyon condensation is given by 

\begin{equation}
\bar \Phi = - \lambda r,~~~ \bar F_{tr} = Q e^{-(2-\epsilon) \lambda r},~~~ \bar T = 0,
~~~ \bar G_{\mu\nu} =
 \left(  \begin{array}{cc} - f & 0  \\
                            0 & f^{-1}   \end{array}   \right)
\end{equation}
with
\begin{equation}
f = 1 -  {M \over \lambda}e^{- 2 \lambda r} + {Q^2 \over 2 \lambda^2(2- \epsilon)}
e^{- (4-\epsilon) \lambda r},
\end{equation}
where $M$ and $Q$ are the mass and electric charge of the black hole, respectively.
Note that from the requirement of $\bar F(r\to \infty) \to 0$ and $f(r\to \infty) \to 1$, we have
the important constraint : $\epsilon <2$.
Hereafter we take $ M= \lambda=\sqrt2$ for convenience.    In the non-extremal black hole,
 from $f=0$ we  obtain 
two roots ($r_{\pm}$)  where $r_{+}(r_{-})$ correspond to the event (Cauchy) horizon.
The extremal black hole may provide a toy model to investigate the late stages 
of Hawking evaporation [13]. This is recovered from the non-extremal black hole in the extremal limit 
($Q \to M: r_- \to r_+ \equiv r_o$). For $\epsilon<2$, the shape of $f$ is always concave.
The multiple root($r=r_o$) is thus obtained when $f(r_o) =0$ and $f^\prime (r_o) = 0$,
in this case the electric charge of the black hole is 
$Q_e^2= 8({2-\epsilon \over 4-\epsilon})^{(4-\epsilon)/2}$. 
Here the prime $(\prime)$ denotes the derivative with respect to $r$.
The extremal horizon is located at 
\begin{equation}
 r_o(\epsilon)= - {1 \over 2 \sqrt 2} \log ({ 4-\epsilon \over 2-\epsilon}).
\end{equation}
The explicit form of the extremal $f$ is 
\begin{equation}
f_e(r,\epsilon) = 1 - e^{- 2 \sqrt 2 r} + {2 \over (2- \epsilon)} \big ( 
{2-\epsilon \over 4-\epsilon} \big)
^{(4-\epsilon)/2}e^{- (4-\epsilon) \sqrt 2 r}.
\end{equation}

Now let us briefly review  the classical aspects of our model.  
We introduce small perturbation fields  around
the background solution as 
\begin{eqnarray}
&&F_{tr} = \bar F_{tr} + {\cal F}_{tr} = \bar F_{tr} [1 - {{\cal F}(r,t) \over Q}],        \\   
&&\Phi = \bar \Phi + \phi(r,t),                       \\  
&&G_{\mu\nu} = \bar G_{\mu\nu} + h_{\mu\nu}  = \bar G_{\mu\nu} [1 - h (r,t)],     \\
&&T = \exp ({\bar \Phi}) [ 0 + \tilde T(r,t) ]. 
\end{eqnarray}
 One has to linearize (2)-(5) in order to obtain the equations governing the perturbations.
However, the classical stability should be based on the physical degrees of freedom. 
It is  thus important to check whether the graviton ($h$),  dilaton ($\phi$), Maxwell
mode (${\cal F}$)  and tachyon ($t$) are  physically propagating modes 
in the 2D charged black hole background. 
According to the counting of the degrees of freedom,
the gravitational field ($h_{\mu\nu}$) in 
$D$-dimensions has $(1/2) D (D -3)$.  For the 4D Schwarzschild black hole,
we obtain two degrees of freedom. These correspond to the Regge-Wheeler mode for odd-parity perturbation
and Zerilli mode for even-parity perturbation [3].  We have $-1$ for $D=2$. This means that in 
two dimensions
the contribution of the graviton is equal and opposite to that of a spinless particle (dilaton).
The graviton-dilaton modes ($h+\phi, h-\phi$) are gauge degrees of freedom and thus are 
nonpropagating modes. 
In addition, the Maxwell field has $D-2$ physical degrees of freedom.
The Maxwell field has no physical degrees of freedom for $D=2$.
Since all these fields are  nonpropagating modes,  equations (2)-(4) are not essential for our study.

On the other hand, the tachyon  is a physically propagating mode.
This is  described by (5) and (13).
Its linearized equation  can be expressed in terms of $\tilde T$ as
\begin{equation}
\nabla^2 \tilde T - [  (\nabla \bar \Phi)^2  - \nabla^2 \bar \Phi - m_0^2] \tilde T = 0.
\end {equation}
From this one finds  
\begin{equation}
f^2 \tilde T''  + ff' \tilde T' - f[ \lambda f' + \lambda^2 f - m_0^2] \tilde T 
 - \partial_t^2 \tilde T = 0.
\end{equation}
To study the classical stability, we should transform the above equation
 into  one-dimensional Schr\"odinger equation by
introducing the tortoise coordinate ($r^*$) 
\begin{equation} 
r\to r^* \equiv g(r).
\end{equation}
Requiring that the coefficient of the linear derivative vanish, one finds the relation
\begin{equation}
g' =  {1 \over f}.
\end{equation}
Assuming $\tilde T( r^*,t ) \sim \tilde T ( r^* ) e^{i\omega t}$, 
one can cast (15) into the Schr\"odinger equation 

\begin{equation}
\{ {d^2 \over dr^{*2}} + \omega^2 - V(r)\} \tilde T = 0,
\end{equation}
where the effective potential $V(r)$  is given by
\begin{equation}
V(r) = f \{(\nabla \bar \Phi)^2  - \nabla^2 \bar \Phi - m_0^2 \} = f\{ \lambda^2 f + \lambda f' - m_0^2\}. 
\end{equation} 
Note that one finds $V(r) = 0$ for a conformally invariant matter $(f_i)$.
For $m_0^2=\lambda^2=2$, it is found that all 2D extremal black holes are classically unstable [5].
Furthermore the outer horizon of 2D non-extremal black hole is stable, while the inner horizon is
unstable.

Now we are in a position to discuss the quantum stability. First of all we have to evaluate 
the one-loop
effective action of the tachyon. From (1) and (13),  the relevant action for the tachyon is rewritten in terms of 
$\tilde T$ as
\begin{equation}
S_{\tilde T} = -{1 \over 2 \pi} \int d^2 x \sqrt{-G} 
   \big \{ (\nabla \tilde T)^2 + [ (\nabla \Phi)^2 - \nabla^2 \bar \Phi -m_0^2] \tilde T^2 \big \}.
\end{equation}
The linearized equation (14) is also derived from the above action.
The coupling of the tachyon to the dilaton is separated as
\begin{equation}
 (\nabla \bar \Phi)^2 - \nabla^2 \bar \Phi -m_0^2 = {\cal Q} +m^2,
\end{equation}
where
\begin{equation}
 {\cal Q}= { V(r) \over f}=(\nabla \bar \Phi)^2 - \nabla^2 \bar \Phi - m_0^2 ;
~~~~~ m^2 = \lambda^2 - m_0^2.
\end{equation}
We are  interested in the massless tachyon, which corresponds to $\lambda^2= m_0^2$.
Quantizing the tachyon in the background of (6) and (7) leads to the one-loop effective action.
Keeping terms quadratic in the classical fields $R$ and ${\cal Q}$, the relevant nonlocal part is 
given by [12]
\begin{equation}
\Gamma_{nloc} = -{1 \over 8 \pi} \int d^2 x \sqrt{-G} 
   \big \{ { 1 \over 12} R {1 \over \nabla^2} R -  {\cal Q} {1 \over \nabla^2} R  
          + {\cal Q} \beta^{(1)} {\cal Q} \big \},
\end{equation}
where 
\begin{equation} 
\beta^{(1)} = { 1 \over \nabla^2} \lim_{ \gamma \to 1} \ln { 1 + \sqrt \gamma \over
 1 - \sqrt \gamma},
\end{equation} 
with
\begin{equation} 
\gamma = { 1 \over  1 - {4 m^2 \over \nabla^2 } }.
\end{equation} 
 The last term is the nonlocal infrared  divergence,
 encounted in the massless limit ($m^2 \to 0$) [14]. Using the $\zeta$-function regularization,
the logarithmic divergences do not appear but one finds the renormalization papameter $\mu$ 
of  the nonlocal infrared divergences. Here we will not consider the last term,
since this depends on the renormalization parameter $\mu$.  For ${\cal Q}= m = 0$,
we get only the first term, which is the well-known result for a conformal matter [8].
For ${\cal Q} \not= 0$ and $ m=0$, the first two-terms contribute the conformal anomaly.
This calculation is easily done in the conformal gauge, rather than the Schwartzschild gauge
in (6). Using the tortoise coordinate $r^*$ in (16), the line element is given by $ds^2= f(-dt^2
+ dr^{*2})$. Comparing this with the general conformal form $( ds^2= e^{2\rho}(-dt^2 + dr^{*2}))$,
one finds  that $\rho = { 1 \over 2} \ln f$. Using this relation, the first two-terms become local
and the trace anomaly is easily computed as
\begin{equation}
\langle T^\alpha_\alpha \rangle 
= - { f'' \over 24 \pi} - {\lambda^2 \over 4 \pi} ( f + 2 f \ln f -\ln f -1 ),
\end{equation}
where the last expression is the new contribution, due to the tachyon coupling to dilaton.
The conservation of stress-energy in the Schwarzschild gauge in (6) leads to
\begin{equation}
\langle T^r_t \rangle = C_1 ;~~  
\langle T^r_r \rangle = { C_2 \over f} + { 1 \over 2f } \int_{r_+}^r dr f' 
\langle T^\alpha_\alpha \rangle,
\end{equation}
where two integration constants $C_1$ and $C_2$ can be determind by considering all information about
the quantum state of the field [15]. For simplicity, we choose $C_1 = C_2 =0$.
 Substituting (26) into (27), then one can perform the integration to find 
\begin{equation}
\langle T^r_r \rangle = - { f'^2 - f'^2(r_+) \over 96 \pi f} - {\lambda^2 \over 8 \pi}(f-1) \ln f.
\end{equation}
Now we turn to the issue of the regularity of stress-energy on the horizon.
Since the line element( $ds^2= -f dt^2 + {1 \over f} dr^2$) is singular on the horizon,
we introduce a freely falling observer. We want to
claculate the stress-energy components in the orthonormal frame attached to
a freely falling observer (FFO).
The basis vectors of the frame are chosen to be the two-velocity $(e_0^\alpha = u^\alpha)$ and
a unit length spacelike vector $(e_1^\alpha = n^\alpha)$ orthogonal to $u^\alpha$.
The components of the tress-energy tensor $\langle T_{ab} \rangle$ in the orthonormal frame are given in 
terms of the coordinate components as
\begin{equation}
\rho = u^\alpha u^\beta \langle T_{\alpha\beta} \rangle = E^2 F(r) - \langle T^r_r \rangle;~~ 
p= n^\alpha n^\beta \langle T_{\alpha\beta} \rangle = E^2 F(r) - \langle T^t_t \rangle,
\end{equation}
where $E$ is the energy per unit mass along the time-like geodesic of FFO and
the quantity $F(r)$ is given by
\begin{equation}
F(r)= {\langle T^r_r \rangle - \langle T^t_t \rangle \over f}.
\end{equation}
The stress-energy tensor is regular on the horizon only if $\langle T^r_r \rangle$, 
$\langle T^t_t \rangle$ and $F(r)$ are each 
separately finite at outer horizon $r=r_+$. Since the strongest possible divergence comes
from $F(r)$, we analyze the divergence structure of this term.
This is computed  near the horizon as
\begin{equation}
\lim_{r \to r_+} F(r) = { 1 \over 48 \pi} \lim_{r \to r_+} { f''' \over f'} + 
{ \lambda^2 \over 4 \pi} \lim_{r \to r_+} (\ln f - { 1 \over f} +1) .
\end{equation}
The first term was discovered by Trivedi for the extremal black holes [10]. From (8) and (9),
one finds $f(r_o)=f'(r_o)=0$ for the extremal black holes and thus one has a weak divergence. 
In additon, here we find  the new divergences from the last term. These are 
the  divergences of the form $\ln f$ and ${1 \over f}$ as $f \to 0$ (near the horizon).
These divergences remain for the non-extremal black holes too, although the form is softened.
Since $f=0$ has a multiple root for extremal black holes, the divergence is stronger than that of 
non-extremal black holes.

In conclusion, we find the new divergences which induce the quantum instability of both the extremal
and non-extremal black holes.

\acknowledgments

This work was supported in part by Nondirected Research Fund, Korea Research Foundation, 1994
 by Korea Science and Engineering Foundation, 94-1400-04-01-3 and 961-0201-005-1.

\end{document}